\documentclass[journal,draftclsnofoot,onecolumn,a4paper]{IEEEtran} 
\usepackage{amsmath}
\usepackage{graphicx}
\usepackage{cite}
\usepackage{amssymb}
\usepackage{multirow}
\usepackage{subfigure}

\title{Performance Analysis of $L$-Branch Scan-and-Wait Combining (SWC) over Arbitrarily Correlated Nakagami-$m$ Fading Channels}
\author{George~C.~Alexandropoulos,~\IEEEmembership{Senior~Member,~IEEE}, P.~Takis~Mathiopoulos,~\IEEEmembership{Senior~Member,~IEEE}, and Pingzhi~Fan,~\IEEEmembership{Fellow,~IEEE}
\thanks{G. C. Alexandropoulos is with Mathematical and Algorithmic Sciences Lab, France Research Center, Huawei Technologies Co. Ltd., 92100 Boulogne-Billancourt, France (e-mail: george.alexandropoulos@huawei.com).}
\thanks{P. T. Mathiopoulos is with Department of Informatics and Telecommunications, National and Kapodistrian University of Athens, 15784 Ilissia, Athens, Greece (e-mail: mathio@di.uoa.gr).}
\thanks{P. Fan is with Institute of Mobile Communications, Southwest Jiaotong University, Chengdu, Sichuan 610031, PR of China (e-mail: p.fan@ieee.org).}}

\begin{document}

\newcounter{my_counter}
\maketitle

\begin{abstract}
The performance of $L$-branch scan-and-wait combining (SWC) reception systems over arbitrarily correlated and not necessarily identically distributed Nakagami-$m$ fading channels is analyzed and evaluated. Firstly, a fast convergent infinite series representation for the SWC output signal-noise ratio (SNR) is presented. This expression is used to obtain analytical expressions in the form of infinite series for the average error probability performance of various modulation schemes for integer values of $m$ as well as the average number of paths estimation and average waiting time (AWT) of $L$-branch SWC receivers for arbitrary values of $m$. The numerically obtained results have shown that the performance expressions converge very fast to their exact analytical values. It was found that the convergence speed depends on the correlation and operating SNR values as well as the Nakagami $m$-parameter. In addition to the analytical results, complementary computer simulated performance evaluation results have been obtained by means of Monte Carlo error counting techniques. The match between these two sets of results has verified the accuracy of the proposed mathematical analysis. Furthermore, it is revealed that, at the expense of a negligible AWT, the average error probability performance of SWC receivers is always superior to that of switched-and-examine combining receivers and in certain cases to that of maximal-ratio combining receivers.
\end{abstract}


\section{Introduction}\label{Sec:Intro}
\IEEEPARstart{S}{patial} diversity techniques play an important role in current mobile broadband systems as efficient means of mitigating channel fading due to multipath and shadowing \cite{B:Proakis, B:Dahlman}. The vast majority of these techniques requires dedicated channel estimation and matched filtering for every diversity branch which, unfortunately, increase their implementation complexity \cite{J:Blanco_79, J:Iskander}. To reduce complexity and processing power consumption, there has been in recent years considerable interest in spatial diversity techniques that utilize only a subset of the available diversity branches, e$.$g$.$ generalized selection combining (GSC) \cite{J:Molisch_Win_Reduced}, minimum selection GSC \cite{C:Kim_MS-GSC}, switched diversity \cite{J:Ko_Opt_2000, J:Yang_Alouini_03, J:Mahaarnichanon_2004, J:Yang_SWC_06, J:Bithas}, minimum estimation and combining GSC \cite{J:Alouini_MEC-GSC} as well as adaptive GSC \cite{J:Kar_AGSD}. It is noted that the high reliability of wireless connections is a crucial requirement especially in emerging technologies optimized for low-cost and low-power consumption \cite{J:Cui_MIMO_WSN, Palomar_2014}, such as low-rate wireless personal area networks and sensor networks \cite{B:WSN}. For such kind of technologies, spatial diversity with a reduced number of radio frequency (RF) chains is an appropriate and promising solution.  

Two of the most popular diversity techniques that utilize a single RF chain are switch-and-stay combining (SSC) \cite{J:Blanco_79, J:Ko_Opt_2000, J:Bithas} and switch-and-examine combining (SEC) \cite{J:Yang_Alouini_03, J:Xiao_Dong_New_Res_on}. In multi-branch SSC, the transmitter/receiver switches to, and stays, with the next available diversity branch regardless of its signal-to-noise ratio (SNR), when the instantaneous SNR on the current branch becomes unacceptable, i$.$e$.$ lower than a preset switching threshold. For the SEC, the combiner first examines the next branch's SNR and switches again only if this SNR is unacceptable. In case where the instantaneous SNR of all available branches is lower than the preset switching threshold, the combiner either uses the last examined branch or switches back to the first branch for the next operation slot. An alternative form of SEC, termed as scan-and-wait combining (SWC), suitable for the sporadic communication of delay-tolerant information over wireless networks, such as wireless ad-hoc and sensor networks, was presented in \cite{J:Yang_SWC_06}. With SWC, if all the available diversity paths fail to meet a predetermined minimum quality requirement, the system waits for a certain time period and restarts the switch-and-examine process. This scanning followed by waiting can then be repeated until a path with acceptable quality is found. It was shown in \cite{J:Yang_SWC_06} that, for a fixed average number of channel estimates per channel access, SWC outperforms SSC and SEC at the expense of a negligible time delay. 

It is well-known that the theoretical gains of multi-antenna systems are degraded in practice due to correlated fading \cite{J:Lombardo_MRD_CorNakagami, J:Zhang_Decom_00, B:Sim_Alou_Book, J:Beaulieu_Inf, J:Mallik_Hybrid, J:Kar_Exponential, J:Kar_Green, J:Tellambura_Rayleigh34, J:Tellambura_Nakagami34, J:Peppas_Nakagami3, J:Reig_ConstantCor, J:Tulino_Correlation, J:Alexandg_TVT}. Correlated fading channels are usually encountered in diversity systems employing antennas which are not sufficiently wide separated from each other, e$.$g$.$ in mobile handsets and indoor base stations. One of the most important fading channel models that incorporates a wide variety of fading environments is the Nakagami-$m$ model \cite{B:Sim_Alou_Book}. In the past, the impact of arbitrarily correlated and not necessarily identically distributed (AC-NNID) Nakagami-$m$ fading on the performance of multi-branch SSC and SEC was investigated in \cite{J:Ko_Opt_2000} and \cite{J:Yang_Alouini_03, J:Alexandg_TVT}, respectively. In \cite{J:Mahaarnichanon_2004}, by deriving an analytical expression for the moment generating function (MGF) of the output SNR of dual-branch SWC receivers over AC-NNID Rayleigh fading channels, the average bit error probability (ABEP) of differential binary phase-shift keying (BPSK) modulation and the average waiting time (AWT) in terms of number of coherence times were evaluated. However, to the best of our knowledge, the performance of $L$-branch SWC receivers with $L>2$ over AC-NNID fading has neither been analyzed nor evaluated so far in the open technical literature. Thus, the purpose of this paper is to fill this gap for the case of AC-NNID Nakagami-$m$ fading channels, and its main contributions are summarized as follows.
\begin{itemize}
	\item The derivation of a generic analytical expression in the form of fast convergent infinite series for the joint probability density function (PDF) of AC-NNID Gamma fading random variables (RVs). This expression is then conveniently used to obtain an analytical expression for the $L$-branch SWC output SNR.
	\item Fast convergent infinite series representations for the AWT and the average number of path estimations (ANPE) of $L$-branch SWC receivers with $L\geq2$ for arbitrary values of the Nakagami $m$-parameter as well as the average error probability of various modulation schemes of the same receivers for integer values of $m$ are derived.	Furthermore, a novel analytical expression for the ANPE of $L$-branch SEC receivers for arbitrary-valued $m$-parameter is presented.
\end{itemize}
To verify the validity of the analytical approach, various numerically evaluated and computer simulation performance evaluation results will be presented and compared. In addition, extensive error probability performance comparisons with $L$-branch SEC and maximal-ratio combining (MRC) receivers are included and discussed.

The reminder of this paper is organized as follows. After this introduction, Section~\ref{Sec:SWD_Statistics} presents the derivation of the analytical expression for the $L$-branch SWC output SNR. In Section~\ref{Sec:Performance}, the various analytical expressions obtained for the performance of SWC receivers are presented. The performance evaluation results are presented and discussed in Section~\ref{Sec:Results}. In Section~\ref{Sec:Conclusion} the conclusions of the paper can be found.

\section{Statistics of the SWC Output SNR} \label{Sec:SWD_Statistics}
A SWC receiver with $L$ antenna branches receiving digitally modulated signals transmitted over a slow varying and frequency nonselective Nakagami-$m$ fading channel is considered. Let $\mathbf{g}^{(L)}=\left[g_1\,g_2\,\cdots\,g_L\right]$ and $\mathbf{g}^{(L)}_{\rm T}=\left[g_{{\rm T}_1}\,g_{{\rm T}_2}\,\cdots\,g_{{\rm T}_L}\right]$ denote the vectors with the instantaneous received SNRs and the predetermined SNR thresholds at the $L$ branches, respectively. For the considered fading model, $g_\ell$ $\forall\,\ell=1,2,\ldots,L$ is a Gamma distributed RV with marginal cumulative distribution function (CDF) given by
\begin{equation}\label{Eq:CDF_Uni_Nak}
F_{g_{\ell}}\left(g\right) = \gamma\left(m,\frac{mg}{\overline{g}_\ell}\right)/\Gamma\left(m\right)
\end{equation} 
where the parameter $m\geq0.5$ relates to the fading severity, $\overline{g}_\ell$ denotes the average received SNR at the $\ell$th branch, $\gamma\left(\cdot,\cdot\right)$ is the lower incomplete Gamma function \cite[eq. (8.350/1)]{B:Gra_Ryz_Book} and $\Gamma(\cdot)$ is the Gamma function \cite[eq. (8.310/1)]{B:Gra_Ryz_Book}. We consider the general case where $g_\ell$'s are arbitrarily correlated with correlation matrix (CM) $\boldsymbol{\Sigma}\in\mathbb{R}^{L\times L}$. This matrix is symmetric, positive definite, and given by $\mathbf{\Sigma}_{k,\ell}\triangleq 1$ $\forall\,k=\ell$, $k=1,2,\ldots,L$, and $\mathbf{\Sigma}_{k,\ell}\triangleq\rho_{k,\ell}$ $\forall\,k\neq\ell$, where $\rho_{k,\ell}\in\left[0,1\right)$ is the correlation coefficient between $g_k$ and $g_\ell$ \cite[eq. (9.195)]{B:Sim_Alou_Book}. With SWC, in the guard period between every two consecutive time slots for data transmission \cite{J:Yang_SWC_06}, the receiver measures $g_1$ of branch $1$ and compares it to $g_{{\rm T}_1}$. If $g_1\geq g_{{\rm T}_1}$, then branch $1$ is selected for information reception in the upcoming time slot. However, if $g_1<g_{{\rm T}_1}$, the receiver switches to branch $2$, measures $g_2$ and compares to $g_{{\rm T}_2}$. This procedure is repeated until either an acceptable branch for information reception is found or all $L$ available branches have been examined without finding any acceptable one, i$.$e$.$ $g_\ell<g_{{\rm T}_\ell}$ $\forall\,\ell$. In the latter case the receiver informs the transmitter through a feedback channel not to transmit during the upcoming time slot and to buffer the input data for a certain waiting period of time. After that waiting period, the system re-initiates the same procedure. Based on this mode of operation, the PDF of the SWC output SNR is given by \cite[eq. (1)]{J:Yang_SWC_06}
\begin{equation}\label{Eq:PDF_SWC_general}
f_{g_{\rm SWC}}\left(g\right) = \frac{\sum_{\ell=1}^{L}f_{g_\ell}^{\rm T}\left(g\right)}{1-F_{\mathbf{g}^{(L)}}
\left(\mathbf{g}^{(L)}_{\rm T}\right)}
\end{equation}
where $F_{\mathbf{g}^{(L)}}\left(\cdot\right)$ denotes the joint CDF of $\mathbf{g}^{(L)}$ and $f_{g_\ell}^{\rm T}\left(\cdot\right)$ is the conditional PDF of the truncated (above $g_{{\rm T}_\ell}$) $g_\ell$ given that $g_1<g_{{\rm T}_1}$, $g_2<g_{{\rm T}_2}$, $\ldots$, $g_{\ell-1}<g_{{\rm T}_{\ell-1}}$. For $\ell=1$ the conditional PDF simplifies to
\begin{subequations}\label{Eq:Truncated_PDF}
\begin{equation}\label{Eq:Truncated_PDF_a}
f_{g_1}^{\rm T}\left(g\right) =
\left\{
\begin{array}{lr}
f_{g_1}\left(g\right),&g\geq g_{{\rm T}_1}\\
0,&{\rm otherwise}\\
\end{array}
\right.
\end{equation}
whereas for $\ell=2,3,\ldots,L$ it can be expressed as
\begin{equation}\label{Eq:Truncated_PDF_b}
f_{g_{\ell}}^{\rm T}\left(g\right) =
\left\{
\begin{array}{lr}
\int_0^{\mathbf{g}^{(\ell-1)}_{\rm T}}
f_{\mathbf{g}^{(\ell)}}
\left(\mathbf{g}^{(\ell-1)},g\right){\rm d}\mathbf{g}^{(\ell-1)},&g>g_{{\rm T}_{\ell}}\\
0,&{\rm otherwise}.\\
\end{array}
\right.
\end{equation}
\end{subequations}
In \eqref{Eq:Truncated_PDF}, $f_{g_{1}}\left(\cdot\right)$ is the marginal PDF of $g_1$ obtained by differentiating \eqref{Eq:CDF_Uni_Nak} \cite[Tab. I]{J:Ko_Opt_2000}, $\int_0^{\mathbf{g}^{(\ell-1)}_{\rm T}}$ denotes multiple integrations $\int_{0}^{g_{{\rm T}_1}}\int_{0}^{g_{{\rm T}_2}}\cdots\int_{0}^{g_{{\rm T}_{\ell-1}}}$ and $f_{\mathbf{g}^{(\ell)}}\left(\cdot\right)$ is the joint PDF of $\mathbf{g}^{(\ell)}$.

Clearly from \eqref{Eq:PDF_SWC_general} and \eqref{Eq:Truncated_PDF}, in order to analyze the performance of $L$-branch SWC receivers over AC-NNID Nakagami-$m$ fading channels, analytical expressions for the joint statistics of $\mathbf{g}^{(L)}$ are needed. A generic analytical expression for the joint CDF of $\mathbf{g}^{(L)}$, that encompasses various cases for $m$, $L$ and the form of $\boldsymbol{\Sigma}$ \cite{J:Beaulieu_Inf, J:Mallik_Hybrid, J:Kar_Exponential, J:Kar_Green, J:Tellambura_Rayleigh34, J:Tellambura_Nakagami34, J:Peppas_Nakagami3, J:Reig_ConstantCor}, was presented in \cite[eq. (6)]{J:Alexandg_TVT}
\begin{equation}\label{Eq:Joint_CDF_Nak}
F_{\mathbf{g}^{\left(L\right)}}\left(\mathbf{g}^{\left(L\right)}\right)=\sum_{\mathbf{i}^{\rm (N)}}^\infty
\mathcal{A}\left(m,\boldsymbol{\Sigma},\mathbf{i}^{(\rm N)}\right)
\prod_{\ell=1}^{L}\gamma\left(\kappa_\ell(m),\xi_\ell(m,\boldsymbol{\Sigma})\frac{g_\ell}{\overline{g}_\ell}\right)
\end{equation}
where $\sum_{\mathbf{i}^{\rm (N)}}^\infty$, with ${\rm N}\in\mathbb{Z}_+^*$, represents the multiple infinite series\footnote{In practice, to numerically evaluate \eqref{Eq:Joint_CDF_Nak}, minimum numbers of terms are selected in the summations leading to a certain accuracy. An upper bound for the resulting truncation error is given by \cite[eq. (7)]{J:Alexandg_TVT}. In addition, \cite{J:Alexandg_TVT} includes results with the minimum numbers of required terms in \eqref{Eq:Joint_CDF_Nak} for convergence to the sixth significant digit for various values of the involved parameters, e$.$g$.$ ${\rm N}$, $m$, $\mathbf{g}^{(L)}$, $\overline{g}_\ell$'s and $\boldsymbol{\Sigma}$.} $\sum_{i_1=0}^\infty\sum_{i_2=0}^\infty\cdots\sum_{i_{\rm N}=0}^\infty$ and $\mathbf{i}^{\rm (N)}=\left[i_1\,i_2\,\cdots\,i_{\rm N}\right]\in\mathbb{Z}_+^{1\times{\rm N}}$. Moreover, as shown in \cite{J:Alexandg_TVT}, ${\rm N}$ and the real-valued scalar functions $\mathcal{A}(\cdot)$, $\kappa_\ell(\cdot)$ and $\xi_\ell(\cdot)$ depend on the form of the CM $\boldsymbol{\Sigma}$ and the fading parameter $m$. For example, considering an exponential CM, i$.$e$.$ $\boldsymbol{\Sigma}_{k,\ell} \triangleq \rho^{|k-\ell|}$ $\forall\,k\neq\ell$ with $|\cdot|$ denoting absolute value, the joint CDF of $\mathbf{g}^{(\ell)}$ is obtained: \textit{i}) by substituting in \eqref{Eq:Joint_CDF_Nak} ${\rm N}=L-1$ and 
\begin{equation}\label{Eq:A_function}
\mathcal{A}\left(m,\boldsymbol{\Sigma},\mathbf{i}^{\rm (N)}\right)=\frac{\det(\mathbf{W})^m}{\Gamma(m)\prod_{\ell=1}^L\mathbf{W}_{\ell,\ell}^{\kappa_\ell(m)}}\prod_{j=1}^{L-1}\frac{\mathbf{W}_{j,j+1}^{2i_j}}{i_j!\Gamma(i_j+m)}
\end{equation}
with $\mathbf{W}_{k,\ell}$ denoting the $(k,\ell)$th element of $\mathbf{W}=(\sqrt{\boldsymbol{\Sigma}})^{-1}$. Also, \textit{ii}) by setting in \eqref{Eq:Joint_CDF_Nak} and \eqref{Eq:A_function} $\kappa_1(m) = i_1+m$, $\kappa_p(m) = i_{p-1}+i_p+m$ for $p=2,3,\ldots,L-1$ and $\kappa_L(m) = i_{L-1}+m$; as well as \textit{iii}) by substituting in \eqref{Eq:Joint_CDF_Nak} $\xi_\ell(m,\boldsymbol{\Sigma}) = m\mathbf{W}_{\ell,\ell}$. Differentiating \eqref{Eq:Joint_CDF_Nak} $L$ times and using \cite[eq. (3.381/1)]{B:Gra_Ryz_Book}, yields the following very generic analytical expression for the joint PDF of $\mathbf{g}^{(L)}$:
\begin{equation}\label{Eq:Joint_PDF_Nak}
f_{\mathbf{g}^{\left(L\right)}}\left(\mathbf{g}^{\left(L\right)}\right)=\sum_{\mathbf{i}^{\rm (N)}}^\infty
\mathcal{A}\left(m,\boldsymbol{\Sigma},\mathbf{i}^{\rm (N)}\right)
\prod_{\ell=1}^{L} \left[\frac{\xi_\ell(m,\boldsymbol{\Sigma})}{\overline{g}_\ell}\right]^{\kappa_\ell(m)}
g_\ell^{\kappa_\ell(m)-1}
\exp\left(-\xi_\ell(m,\boldsymbol{\Sigma})\frac{g_\ell}{\overline{g}_\ell}\right).
\end{equation}

Substituting $f_{g_{1}}\left(g\right)$ in \eqref{Eq:Truncated_PDF_a} for $g\geq g_{{\rm T}_1}$ as well as \eqref{Eq:Joint_PDF_Nak} in \eqref{Eq:Truncated_PDF_b} for $g>g_{{\rm T}_\ell}$ $\forall\,\ell\geq2$ and using \cite[eq. (3.381/1)]{B:Gra_Ryz_Book} to solve the resulting multiple integrals, an analytical expression for the numerator of \eqref{Eq:PDF_SWC_general} is obtained. Using this expression and substituting \eqref{Eq:Joint_CDF_Nak} in the denominator of \eqref{Eq:PDF_SWC_general}, a general analytical expression for the PDF of the SWC output SNR over AC-NNID Nakagami-$m$ fading channels is given by  
\begin{equation}\label{Eq:SWD_Output_NAK}
\begin{split}
&f_{g_{\rm SWC}}\left(g\right) = \left[1-\sum_{\mathbf{i}^{({\rm N}_L)}}^\infty
\mathcal{A}\left(m,\boldsymbol{\Sigma}^{(L)},\mathbf{i}^{({\rm N}_L)}\right)
\prod_{\ell=1}^{L}\gamma\left(\kappa_\ell(m),\xi_\ell(m,\boldsymbol{\Sigma}^{(L)})\frac{g_{{\rm T}_\ell}}{\overline{g}_\ell}\right)\right]^{-1}
\\&\times\left\{\left(\frac{m}{\overline{g}_1}\right)^{m}\frac{g^{m-1}}{\Gamma(m)}\exp\left(-\frac{mg}{\overline{g}_1}\right)
+\sum_{\ell=2}^L \sum_{\mathbf{i}^{({\rm N}_\ell)}}^\infty \mathcal{A}\left(m,\boldsymbol{\Sigma}^{(\ell)},\mathbf{i}^{({\rm N}_\ell)}\right)
\left[\prod_{j=1}^{\ell-1}\gamma\left(\kappa_j(m),\xi_j(m,\boldsymbol{\Sigma}^{(\ell)})\frac{g_{{\rm T}_j}}{\overline{g}_j}\right)\right]
\right.
\\&\left.
\times\left[\frac{\xi_\ell(m,\boldsymbol{\Sigma}^{(\ell)})}{\overline{g}_\ell}\right]^{\kappa_\ell(m)}g^{\kappa_\ell(m)-1}
\exp\left(-\xi_\ell(m,\boldsymbol{\Sigma}^{(\ell)})\frac{g}{\overline{g}_\ell}\right)\right\}.
\end{split}
\end{equation}

This expression is valid for $g\geq g_{{\rm T}_1}$ and $g>g_{{\rm T}_\ell}$ $\forall\,\ell\geq2$, and ${\rm N}_\ell\in\mathbb{Z}_+^*$ $\forall\,\ell\geq2$ depends on the form of the CM $\boldsymbol{\Sigma}^{(\ell)}$ of the instantaneous SNRs of the $\ell$ diversity branches \cite{J:Alexandg_TVT}. For example, ${\rm N}_2=1$ for an arbitrary CM $\boldsymbol{\Sigma}^{(2)}$, while for $\ell\geq3$ and exponentially correlated $\boldsymbol{\Sigma}^{(\ell)}$, ${\rm N}_\ell=\ell-1$. To evaluate \eqref{Eq:SWD_Output_NAK}, $\sum_{\ell=2}^L{\rm N}_\ell$ infinite series in the numerator and ${\rm N}_L$ in the denominator need to be evaluated, for which minimum numbers of terms are selected in practice. It is also noted that \eqref{Eq:SWD_Output_NAK} can be easily modified to account for the cases where $g<g_{{\rm T}_1}$ and/or $g\leq g_{{\rm T}_\ell}$ for any $\ell\geq2$. In such cases, the first and/or the $\ell$th multiple infinite series term in the numerator of \eqref{Eq:SWD_Output_NAK} should be set to $0$.   

\section{Performance Analysis} \label{Sec:Performance}
The previously derived formulas will be used to obtain analytical expressions for the performance of $L$-branch SWC receivers, in terms of AWT, ANPE and average error probability, over AC-NNID Nakagami-$m$ fading channels. \setcounter{equation}{7}

\subsection{AWT and ANPE}\label{sub:AWT_SWD_NAK}
To quantify the complexity and required processing power consumption as well as the delay of SWC receivers, the following performance metrics are considered \cite{J:Yang_SWC_06}: \emph{i}) The ANPE, $\overline{N}_{\rm e}$, before channel access; and \emph{ii}) The AWT, $\overline{N}_{\rm c}$, in terms of the number of coherence times that the system has to wait before an acceptable path is found and transmission occurs. By substituting \eqref{Eq:CDF_Uni_Nak} and \eqref{Eq:Joint_CDF_Nak} into \cite[eq. (28)]{J:Yang_SWC_06}, the following general analytical expression for $\overline{N}_{\rm e}$ can be obtained
\begin{equation}\label{Eq:ANPE_Nak}
\overline{N}_{\rm e} = \overline{N}_{\rm SEC}\left[1-F_{\mathbf{g}^{\left(L\right)}}\left(\mathbf{g}^{(L)}_{\rm T}\right)\right]^{-1}
\end{equation}
where $\overline{N}_{\rm SEC}$ represents ANPE of SEC, which is given by
\begin{equation}\label{Eq:ANPE_SED_Nak_def}
\overline{N}_{\rm SEC} = \sum_{\ell=1}^{L}\ell{\rm Pr}\left[N_{\rm SEC}=\ell\right].
\end{equation}
In \eqref{Eq:ANPE_SED_Nak_def}, ${\rm Pr}\left[\cdot\right]$ denotes probability and $N_{\rm SEC}$ is a discrete RV representing the number of path estimations for SEC. By using \eqref{Eq:CDF_Uni_Nak} and \eqref{Eq:Joint_CDF_Nak}, and after some straightforward algebraic manipulations, $\overline{N}_{\rm SEC}$ can be expressed as
\begin{equation}\label{Eq:ANPE_SED_Nak}
\overline{N}_{\rm SEC} =  1+\sum_{\ell=1}^{L-1}F_{\mathbf{g}^{\left(\ell\right)}}\left(\mathbf{g}^{(\ell)}_{\rm T}\right).
\end{equation}
Furthermore, by substituting \eqref{Eq:Joint_CDF_Nak} into \cite[eq. (16)]{J:Yang_SWC_06}, a generic analytical expression for $\overline{N}_{\rm c}$ is derived as
\begin{equation}\label{Eq:AWT_Nak}
\overline{N}_{\rm c} = F_{\mathbf{g}^{\left(L\right)}}\left(\mathbf{g}^{(L)}_{\rm T}\right)\left[1-F_{\mathbf{g}^{\left(L\right)}}\left(\mathbf{g}^{(L)}_{\rm T}\right)\right]^{-1}.
\end{equation}

\subsection{Average Error Probability}\label{sub:ABEP_SWD_NAK}
The average error probability performance of digital modulation schemes of $L$-branch SWC receivers over fading channels can be evaluated using
\begin{equation}\label{Eq:ASEP_definition}
\overline{P}_{\rm e}=\int_{0}^{\infty} P_{\rm cp}\left(g\right)f_{g_{\rm SWC}}\left(g\right){\rm d}g
\end{equation}
where $P_{\rm cp}(\cdot)$ depends on the modulation scheme \cite[Chap$.$ 5]{B:Proakis}, \cite[Chap$.$ 8]{B:Sim_Alou_Book}. For a great variety of such schemes, $P_{\rm cp}(g)=AQ(\sqrt{Bg})$ where $A,B\in\mathbb{R}_+^*$ are constants and $Q(\cdot)$ denotes the Gaussian $Q$-function \cite[eq. (4.1)]{B:Sim_Alou_Book}. For example, $A=1$ and $B=2$ leads to binary phase-shift keying (BPSK) whereas, $A=2(1-1/M)$ and $B=6/(M^2-1)$ to $M$-ary pulse amplitude modulation ($M$-PAM). In addition, a tight approximation for ABEP of rectangular $M$-ary pulse quadrature modulation ($M$-QAM) can be obtained with $A=4(1-1/\sqrt{M})$ and $B=3/(M-1)$. By substituting \eqref{Eq:PDF_SWC_general} into \eqref{Eq:ASEP_definition}, the average error probability of various modulation schemes of $L$-branch SWC receivers over fading channels can be expressed as 
\begin{equation}\label{Eq:ASEP_definition_SWD}
\overline{P}_{\rm e} = \frac{A}{1-F_{\mathbf{g}^{(L)}}
\left[\mathbf{g}^{(L)}_{\rm T}\right]}\sum_{\ell=1}^{L}\int_{g_{{\rm T}_\ell}}^{\infty}Q(\sqrt{Bg})f_{g_\ell}^{\rm T}\left(g\right){\rm d}g.
\end{equation}

To evaluate the average error probability for AC-NNID Nakagami-$m$ fading channels, we first substitute $f_{g_{1}}\left(g\right)$ and \eqref{Eq:Joint_PDF_Nak} into \eqref{Eq:Truncated_PDF_a} and \eqref{Eq:Truncated_PDF_b}, respectively, and then into \eqref{Eq:ASEP_definition_SWD}. To this end, integrals of the form $I_\ell(b,c)=\int_{g_{{\rm T}_\ell}}^{\infty} Q(\sqrt{Bg})g^{b-1}\exp\left(-cg\right){\rm d}g$ $\forall$ $\ell=1,2,\ldots,L$, with $b\geq0.5$ and $c>0$, need to be evaluated. As will be shown in the sequel, parameter $b$ depends on $m$ whereas, parameter $c$ is a function of $m$, the average received SNRs and the correlation coefficients. Similar to \cite[Sec$.$ III.A]{J:Xiao_Dong_New_Res_on} and using integration by parts, $I_\ell(b,c)$ can be rewritten as  
\begin{equation}\label{Eq:Integral}
I_\ell(b,c) = c^{-b}\left[Q\left(\sqrt{Bg_{{\rm T}_\ell}}\right)\Gamma\left(b,g_{{\rm T}_\ell}c\right)-\frac{Y_\ell(b,c)}{\sqrt{2\pi}}\right]
\end{equation}
where $\Gamma\left(\cdot,\cdot\right)$ is the upper incomplete Gamma function \cite[eq. (8.350/2)]{B:Gra_Ryz_Book} and $Y_\ell(b,c)$ is given by
\begin{equation}\label{Eq:Integral_Y}
Y_\ell(b,c) = \int_{\sqrt{Bg_{{\rm T}_\ell}}}^{\infty} \exp\left(-0.5g^2\right)\Gamma\left(b,\frac{cg^2}{B}\right){\rm d}g.
\end{equation}
A closed-form solution for the integral in \eqref{Eq:Integral_Y} can be obtained for integer-valued $b\geq1$ by using \cite[eq. (8.352/7)]{B:Gra_Ryz_Book} and \cite[eq. (8.381/3)]{B:Gra_Ryz_Book}, as
\begin{equation}\label{Eq:Integral_Solution}
Y_\ell(b,c) = 0.5\sqrt{B}(b-1)!\sum_{n=0}^{b-1}\frac{c^n\Gamma\left(n+0.5,\left(0.5B+c\right)g_{{\rm T}_\ell}\right)}{n!\left(0.5B+c\right)^{n+0.5}}.
\end{equation}
Finally, by substituting \eqref{Eq:Integral_Solution} into \eqref{Eq:Integral} and then in \eqref{Eq:ASEP_definition_SWD}, yields a generic analytical expression for the average error probability performance of various modulation schemes over AC-NNID Nakagami-$m$ fading with integer $m$, given by  
\begin{equation}\label{Eq:ASEP_NAK}
\begin{split}
\overline{P}_{\rm e} =& A\left[1-\sum_{\mathbf{i}^{({\rm N}_L)}}^\infty
\mathcal{A}\left(m,\boldsymbol{\Sigma}^{(L)},\mathbf{i}^{({\rm N}_L)}\right)
\prod_{\ell=1}^{L}\gamma\left(\kappa_\ell(m),\xi_\ell(m,\boldsymbol{\Sigma}^{(L)})\frac{g_{{\rm T}_\ell}}{\overline{g}_\ell}\right)\right]^{-1}\left\{\frac{\left(m/\overline{g}_1\right)^{m}}{\Gamma(m)}I_1\left(m,\frac{m}{\overline{g}_1}\right)\right.
\\&\left.+\sum_{\ell=2}^L \sum_{\mathbf{i}^{({\rm N}_\ell)}}^\infty \mathcal{A}\left(m,\boldsymbol{\Sigma}^{(\ell)},\mathbf{i}^{({\rm N}_\ell)}\right)
\left[\prod_{j=1}^{\ell-1}\gamma\left(\kappa_j(m),\xi_j(m,\boldsymbol{\Sigma}^{(\ell)})\frac{g_{{\rm T}_j}}{\overline{g}_j}\right)\right]
\left[\frac{\xi_\ell(m,\boldsymbol{\Sigma}^{(\ell)})}{\overline{g}_\ell}\right]^{\kappa_\ell(m)}\right.
\\&\left.\times I_\ell\left(\kappa_\ell(m),\frac{\xi_\ell(m,\boldsymbol{\Sigma}^{(\ell)})}{\overline{g}_\ell}\right)\right\}.
\end{split}
\end{equation}
For the special case of dual-branch SWC receivers operating over AC-NNID Rayleigh channels, by setting $L=2$ and $m=1$ in \eqref{Eq:ASEP_NAK} and then by substituting using \cite[eqs. (11) and (12)]{J:Alexandg_TVT}\footnote{A typo exists in the numerator in the right-hand side of the equality in \cite[eq. (11b)]{J:Alexandg_TVT}, where $\kappa_1$ needs to be replaced by $k_1$.} the resulting parameters as ${\rm N}_2=1$, $\mathcal{A}(1,\boldsymbol{\Sigma}^{(2)},i_1)=\left(1-\sqrt{\rho}\right)\rho^{i_1/2}/[i_1!\Gamma(i_1+1)]$ and $\xi_\ell(1,\boldsymbol{\Sigma}^{(2)})=\left(1-\sqrt{\rho}\right)$ for $\ell=1$ and $2$, the following analytical expression for the error probability is deduced
\begin{equation}\label{Eq:ASEP_Ray_L2}
\begin{split}
\overline{P}_{\rm e} =& A\left[1-
\left(1-\sqrt{\rho}\right)\sum_{i_1=0}^\infty\frac{\rho^{i_1/2}}{i_1!\Gamma(i_1+1)}
\prod_{\ell=1}^{2}\gamma\left(i_1+1,\frac{g_{{\rm T}_\ell}}{\left(1-\sqrt{\rho}\right)\overline{g}_\ell}\right)\right]^{-1}
\left\{\frac{1}{\overline{g}_1}I_1\left(1,\frac{1}{\overline{g}_1}\right)\right.
\\&\left.+\left(1-\sqrt{\rho}\right)\sum_{i_1=0}^\infty\frac{\rho^{i_1/2}}{i_1!\Gamma(i_1+1)}
\gamma\left(i_1+1,\frac{g_{{\rm T}_1}}{\left(1-\sqrt{\rho}\right)\overline{g}_1}\right)
\left[\frac{1}{\left(1-\sqrt{\rho}\right)\overline{g}_2}\right]^{i_1+1}\right.
\\&\left.\times I_2\left(i_1+1,\frac{1}{\left(1-\sqrt{\rho}\right)\overline{g}_2}\right)\right\}.
\end{split}
\end{equation}
For this special case, the average error probability of various modulation schemes can also be evaluated using the MGF-based approach \cite[Chap. 1]{B:Sim_Alou_Book} with the MGF expression given by \cite[eq. (7)]{J:Mahaarnichanon_2004}. Although the latter expression was derived using the bivariate Rayleigh CDF given by \cite[eq. (6.5)]{B:Sim_Alou_Book} while in obtaining \eqref{Eq:ASEP_Ray_L2} the alternative infinite series representation \cite[eq. (4)]{J:Beaulieu_Inf} for this CDF was used, it will be shown in the following section, where the performance evaluation results will be presented, that both approaches yield identical average error probability performance.  

\section{Performance Evaluation Results} \label{Sec:Results}
In this section, the analytical expressions \eqref{Eq:ANPE_Nak}, \eqref{Eq:AWT_Nak}, \eqref{Eq:ASEP_NAK} and \eqref{Eq:ASEP_Ray_L2} have been used to evaluate the performance of $L$-branch SWC receivers over AC-NNID Nakagami-$m$ fading channels. Furthermore, and similar to \cite{J:Yang_SWC_06}, we have compared the average error performance of various modulation schemes of the considered receivers with that of SEC and MRC receivers. To this end, the analytical expressions \cite[eq. (21)]{J:Alexandg_TVT} and \cite[eq. (11)]{J:Lombardo_MRD_CorNakagami} for the average symbol error probability (ASEP) performance of SEC and MRC, respectively, have been also evaluated. To verify the correctness of the proposed analysis, equivalent performance evaluation results obtained by means of Monte Carlo simulations will be also presented. For this, and in order to generate $L$ AC-NNID Gamma RVs with CM $\boldsymbol{\Sigma}$, the decomposition technique of \cite{J:Zhang_Decom_00} has been used. Accordingly, each of the $L$ AC-NNID Gamma RVs is obtained as a sum of $2m$ AC-NNID Gaussian RVs with CM $\sqrt{\boldsymbol{\Sigma}}$. Without loss of generality, for the performance results obtained we have considered: \textit{i}) independent and identically distributed (IID) fading with $\rho_{k,\ell}=10^{-4}$ $\forall\,k\neq\ell$ as well as $\overline{g}_\ell=\overline{g}_1$ and $g_{{\rm T}_\ell}=g_{{\rm T}_1}$ $\forall\,\ell>1$; and \textit{ii}) AC-NNID fading characterized by exponential CMs with $\rho=0.9$ and the exponential power decaying profiles $\overline{g}_\ell=\overline{g}_1\exp\left(-\delta\left(\ell-1\right)\right)$ and $\overline{g}_{{\rm T}_\ell}=\overline{g}_{{\rm T}_1}\exp\left(-\delta\left(\ell-1\right)\right)$ $\forall\,\ell$, where $\delta$ is the power decaying factor. 
\begin{table*}[!t]
\renewcommand{\arraystretch}{1}
\caption{The Minimum Number of Terms, $N_{\rm min}$, Required in Each of the Series in \eqref{Eq:ANPE_Nak}, \eqref{Eq:AWT_Nak}, \eqref{Eq:ASEP_NAK} and \eqref{Eq:ASEP_Ray_L2} for the Convergence of the Performance Evaluation Results Shown in Figs.~\ref{Fig:ASEP_SWD_SED}, \ref{Fig:ASEP_SWD_MRD} and \ref{Fig:ASEP_SWD_MRD_MQAM} for $\rho=0.9$ with error less or equal to $10^{-6}$.}
\label{Tab:Convergence_ASEP}\centering
\begin{tabular}{|c||c|c||c|c||c|c||}
\hline\hline
& \multicolumn{2}{c||}{Fig.~\ref{Fig:ASEP_SWD_SED}} & \multicolumn{2}{c||}{Fig.~\ref{Fig:ASEP_SWD_MRD}} & \multicolumn{2}{c||}{Fig.~\ref{Fig:ASEP_SWD_MRD_MQAM}}\\[0.5ex]
$\overline{g}_1$ (dB) & $L=3$, $m=1$ & $L=3$, $m=3$ & $L=2$, $m=1$ & $L=5$, $m=1$ & $L=3$, $m=2$ & $L=5$, $m=2$\\
\hline 0 &10 & 21 & 10 & 14 & 20 & 22\\
\hline 2.5 &8 & 18 & 7 & 7 & 20 & 22\\
\hline 5 &7 & 18 & 7 & 7 & 18 & 20\\
\hline 7.5 &7 & 15 & 6 & 5 & 16 & 17\\
\hline 10 &6 & 11 & 4 & 4 & 15 & 13\\
\hline 12.5 &5 & 11 & 3 & 4 & 12 & 10\\
\hline 15 &4 & 8 & 2 & 2 & 8 & 9\\
\hline\hline
\end{tabular}
\end{table*}

\begin{figure}[!t]
\centering
\includegraphics[keepaspectratio,width=4.5in]{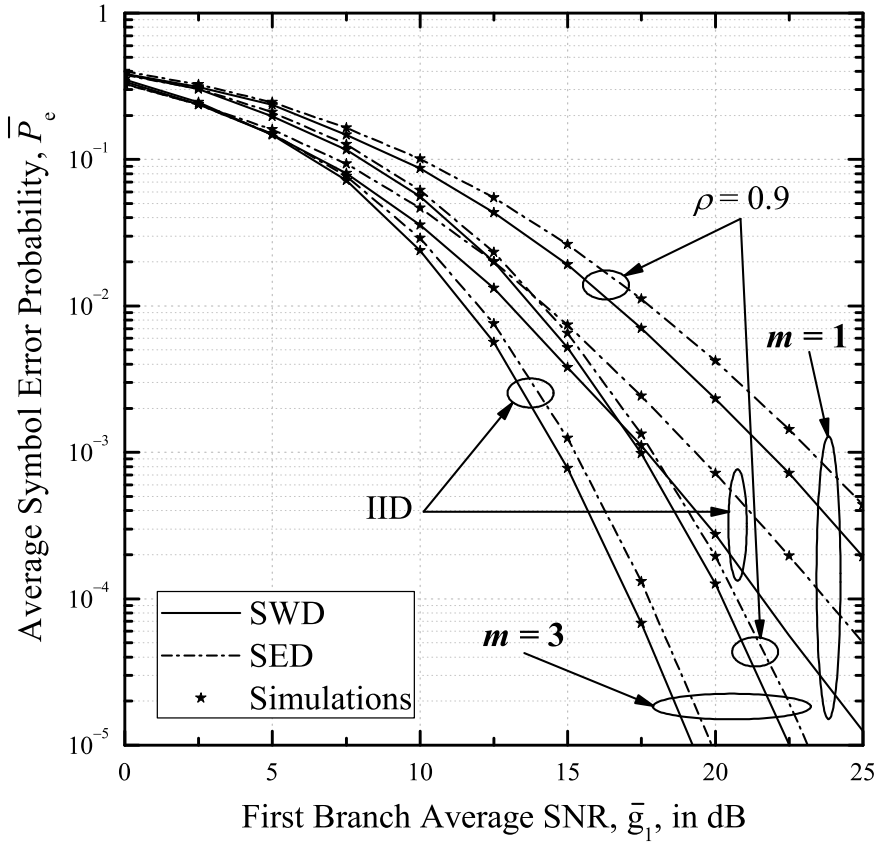}
\caption{ASEP, $\overline{P}_{\rm e}$, of $4$-PAM versus the first branch average SNR, $\overline{g}_1$, for 3-branch diversity receivers over AC-NNID Nakagami-$m$ fading channels.} \label{Fig:ASEP_SWD_SED}
\end{figure}
\begin{figure}[!t]
\centering
\includegraphics[keepaspectratio,width=4.5in]{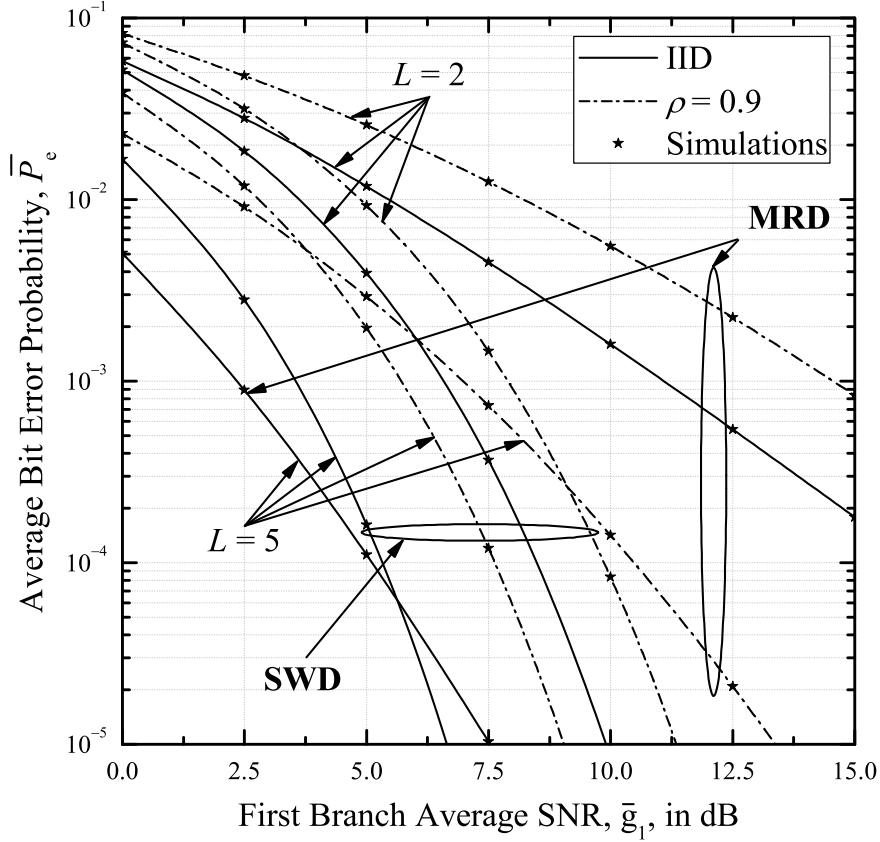}
\caption{ABEP, $\overline{P}_{\rm e}$, of BPSK modulation versus the first branch average SNR, $\overline{g}_1$, for $L$-branch diversity receivers over AC-NNID Rayleigh fading channels.} \label{Fig:ASEP_SWD_MRD}
\end{figure}
\begin{figure}[!t]
\centering
\includegraphics[keepaspectratio,width=4.5in]{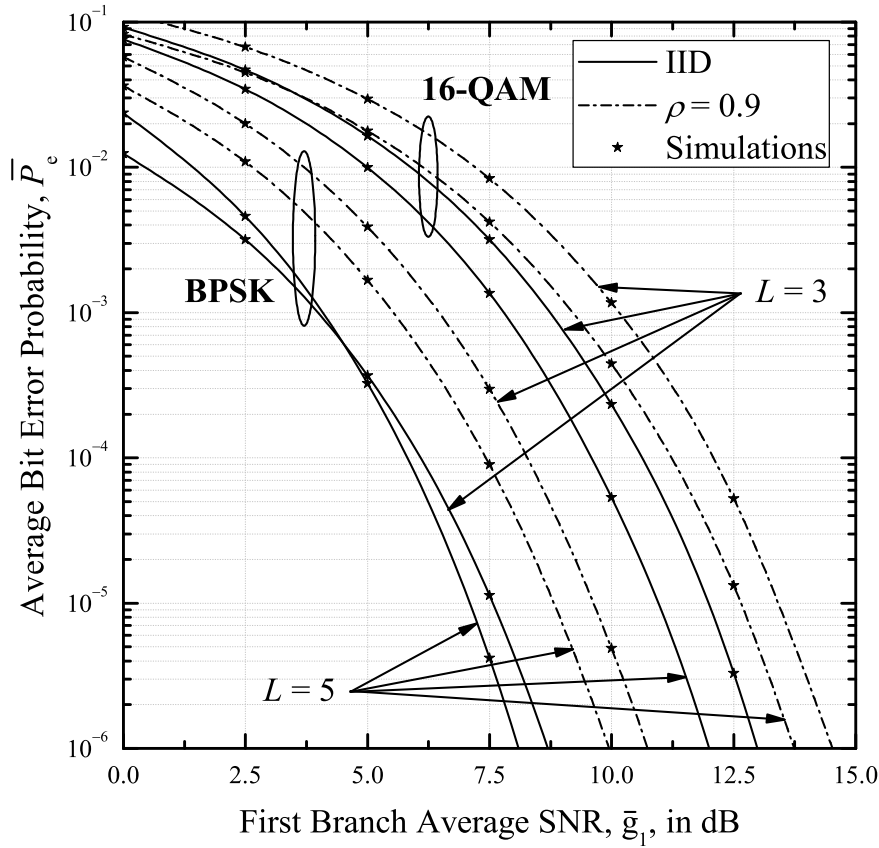}
\caption{ABEP, $\overline{P}_{\rm e}$, of BPSK and $16$-QAM modulations versus the first branch average SNR, $\overline{g}_1$, for $L$-branch SWC receivers over AC-NNID Nakagami-$m$ fading channels with $m=2$.} \label{Fig:ASEP_SWD_MRD_MQAM}  
\end{figure}
\begin{figure}[!t]
\centering
\includegraphics[keepaspectratio,width=4.5in]{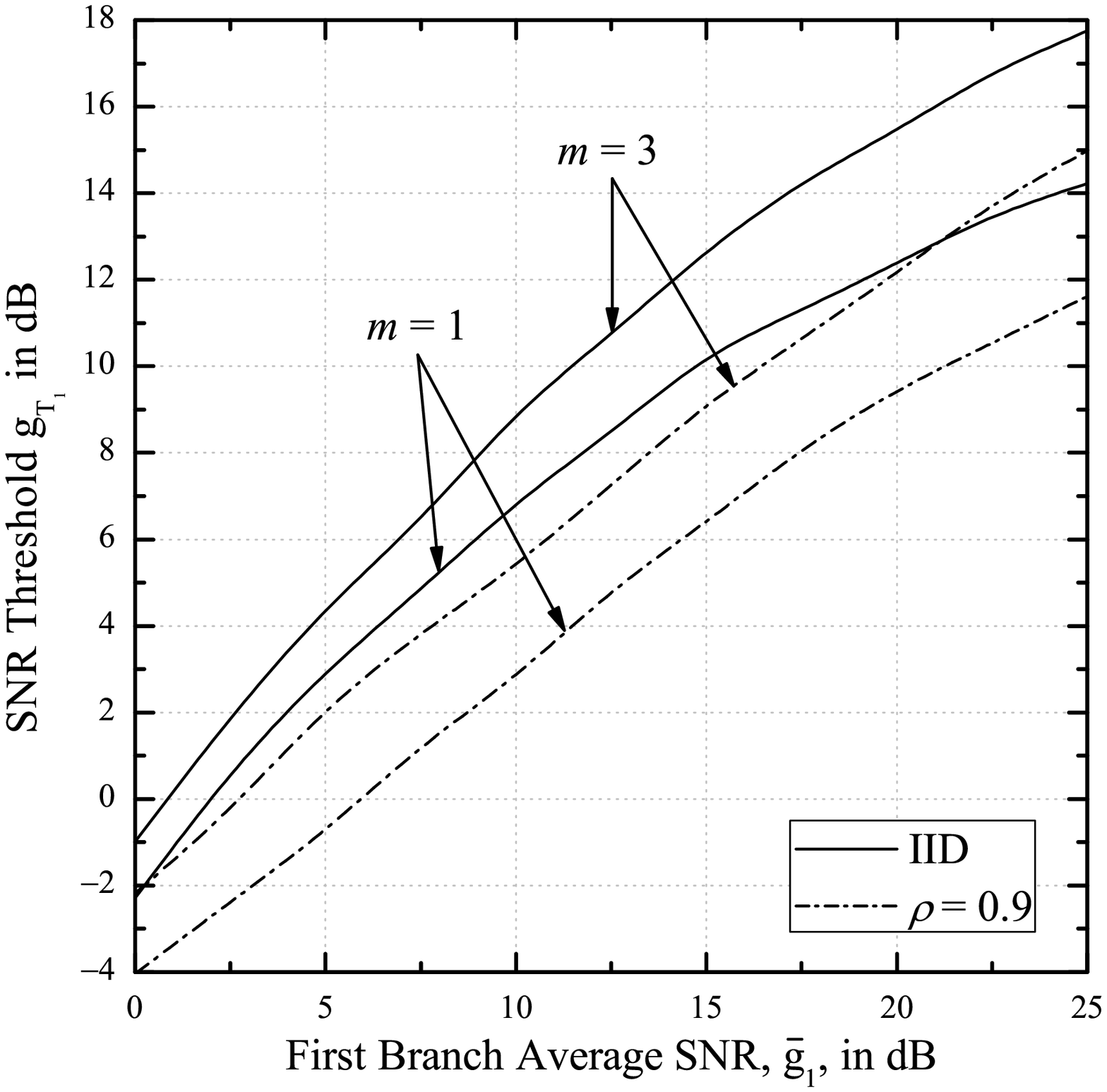}
\caption{SNR threshold $g_{{\rm T}_1}$ versus the first branch average SNR, $\overline{g}_1$, for the $3$-branch SWC receivers in Fig$.$~\ref{Fig:ASEP_SWD_SED}.} \label{Fig:Optimum_gT_SWD_SED}
\end{figure}
\begin{figure}[!t]
\centering
\includegraphics[keepaspectratio,width=4.5in]{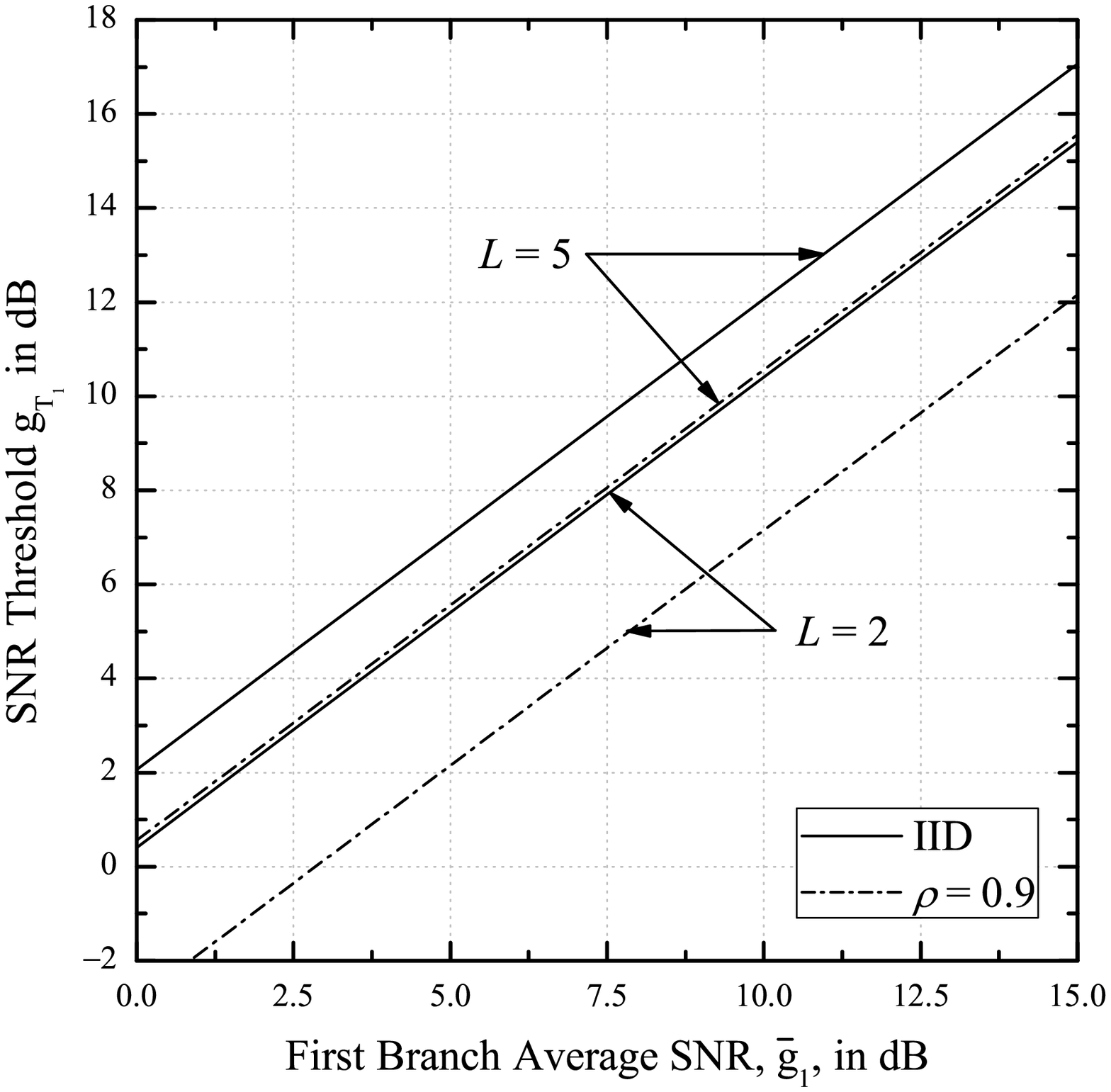}
\caption{SNR threshold $g_{{\rm T}_1}$ versus the first branch average SNR, $\overline{g}_1$, for the $L$-branch SWC receivers in Fig$.$~\ref{Fig:ASEP_SWD_MRD}.} \label{Fig:Optimum_gT_SWD_MRD}
\end{figure}
\begin{figure}[!t]
\centering
\includegraphics[keepaspectratio,width=4.5in]{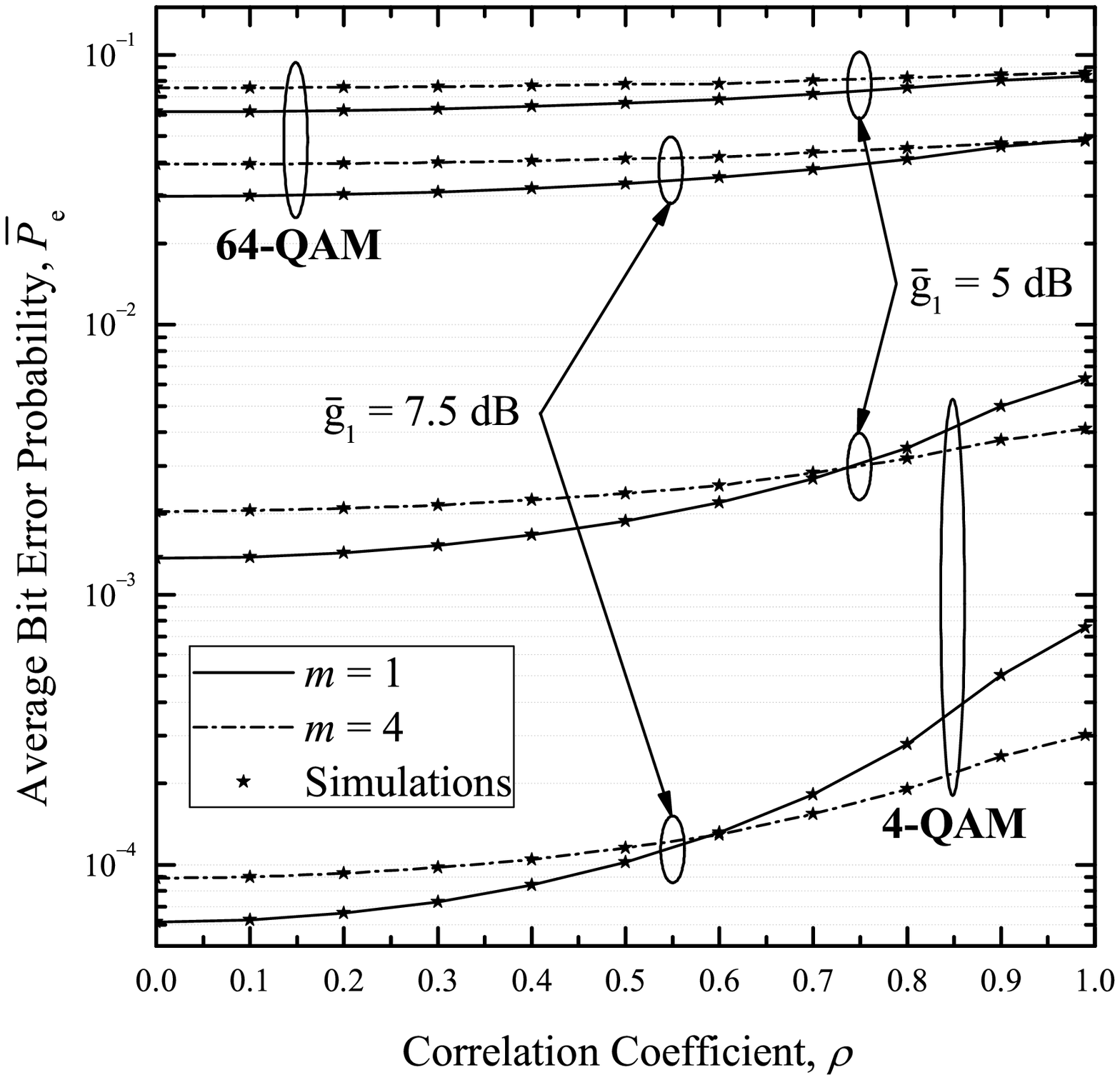}
\caption{ABEP, $\overline{P}_{\rm e}$, of $4$- and $64$-QAM modulations versus the correlation coefficient, $\rho$, for $3$-branch SWC receivers over AC-NNID Nakagami-$m$ fading channels.} \label{Fig:ASEP_vs_rho}
\end{figure}

In Figs$.$~\ref{Fig:ASEP_SWD_SED} and~\ref{Fig:ASEP_SWD_MRD}, the average error performance of $L$-branch SWC receivers is depicted as a function of the first branch average SNR, $\overline{g}_1$, and compared with the performance of SEC and MRC reception, respectively. Similar to \cite{J:Yang_SWC_06}, and for a fair comparison, in Fig$.$~\ref{Fig:ASEP_SWD_SED} $\overline{N}_{\rm e}$ was set equal to $\overline{N}_{\rm SEC}$ and in Fig.~\ref{Fig:ASEP_SWD_MRD} it was set equal $L$, i$.$e$.$ equal to the ANPE of MRC. In particular, in Fig$.$~\ref{Fig:ASEP_SWD_SED} for $L=3$, Nakagami-$m$ fading with $m=1$ and $3$, $\delta=0$ and $4$-PAM, we have used the ASEP expression \cite[eq. (21)]{J:Alexandg_TVT} for each $\overline{g}_1$ value to compute the optimum SNR threshold $g_{{\rm T}_1}^*$ minimizing SEC's ASEP, as described in \cite[Sec. III.B]{J:Alexandg_TVT}. This $g_{{\rm T}_1}^*$ was next substituted into \eqref{Eq:ANPE_SED_Nak} to evaluate $\overline{N}_{\rm SEC}$ that was then used in \eqref{Eq:ANPE_Nak} to solve for $g_{{\rm T}_1}$ of SWC. The computed $g_{{\rm T}_1}$ values for the $3$-branch SWC receivers the performance of which is shown in Fig$.$~\ref{Fig:ASEP_SWD_SED} as a function of $\overline{g}_1$, are depicted in Fig$.$~\ref{Fig:Optimum_gT_SWD_SED}. These values were utilized in \eqref{Eq:ASEP_NAK} to evaluate ASEP of SWC. In Fig$.$~\ref{Fig:ASEP_SWD_MRD}, for the SWC's ABEP curves for $L=2$ and $5$, Rayleigh fading (i$.$e$.$ Nakagami with $m=1$), $\delta=0.1$ and BPSK modulation, \eqref{Eq:ANPE_Nak} was first set to $2$ and $5$ for each $\overline{g}_1$ value to solve for $g_{{\rm T}_1}$ needed in \eqref{Eq:ASEP_Ray_L2} for $2$-branch and in \eqref{Eq:ASEP_NAK} for $5$-branch SWC reception, respectively. 
The resulting $g_{{\rm T}_1}$ values are plotted in Fig$.$~\ref{Fig:Optimum_gT_SWD_MRD} versus $\overline{g}_1$. In Fig$.$~\ref{Fig:ASEP_SWD_MRD}, the ASEP with BPSK for MRC was evaluated using the MGF-based approach \cite[Chap. 1]{B:Sim_Alou_Book} and \cite[eq. (11)]{J:Lombardo_MRD_CorNakagami}. Similar to the approach for deriving the performance curves in Fig$.$~\ref{Fig:ASEP_SWD_MRD}, in Figs$.$~\ref{Fig:ASEP_SWD_MRD_MQAM} and~\ref{Fig:ASEP_vs_rho} we have set \eqref{Eq:ANPE_Nak} equal to $3$ and $5$, i$.$e$.$ $\overline{N}_{\rm e}$ was set equal to the ANPE of $3$- and $5$-branch MRC, respectively, to obtain ABEP of BPSK and $M$-QAM modulation schemes of $3$- and $5$-branch SWC receivers for various values of $m$ and $\rho$. We have selected $\delta=0.1$ only for the correlated case in Fig$.$~\ref{Fig:ASEP_SWD_MRD_MQAM} and for every value of $\rho$ in Fig$.~$\ref{Fig:ASEP_vs_rho}. 
For the numerical evaluation of \eqref{Eq:ANPE_Nak}, \eqref{Eq:AWT_Nak}, \eqref{Eq:ASEP_NAK} and \eqref{Eq:ASEP_Ray_L2} in the performance results shown in all figures, the minimum number of terms, $N_{\rm min}$, was used in all involved series of each multiple series term so that they converge with error less or equal to $10^{-6}$. The maximum value of $N_{\rm min}$ was $2$ in all results for IID fading whereas, Table~\ref{Tab:Convergence_ASEP} summarizes the required number of $N_{\rm min}$ to achieve this accuracy for the considered AC-NNID fading cases. As clearly shown in this table, $N_{\rm min}$ values are quite low and increase as $\overline{g}_1$ decreases and/or $m$ and/or $L$ increase.

The results in Figs$.$~\ref{Fig:ASEP_SWD_SED}--\ref{Fig:ASEP_SWD_MRD_MQAM} illustrate that error performance for all considered receivers improves with increasing $\overline{g}_1$ and/or $m$ and/or decreasing $\rho$. Furthermore, as shown in Fig$.$~\ref{Fig:ASEP_SWD_SED}, SWC outperforms SEC and this superiority is more pronounced when $m$ and/or $\rho$ and/or $\overline{g}_1$ decrease. From the SWC's ASEP curves in Fig$.$~\ref{Fig:ASEP_SWD_SED}, it is also evident from Fig$.$~\ref{Fig:Optimum_gT_SWD_SED} that increasing $\overline{g}_1$ and/or $m$ and/or decreasing $\rho$ results in increasing $g_{{\rm T}_1}$. In addition, it can be observed that AWT increases as $\overline{g}_1$ and/or $\rho$ decrease and/or $m$ increases. For example, considering IID fading channels, and for $\overline{g}_1=5$ dB, $\overline{N}_{\rm c}=0.107$ and $0.121$ for $m=1$ and $3$, respectively, while for the AC-NNID case $\overline{N}_{\rm c}=0.060$ and $0.089$ for $m=1$ and $3$, respectively. Additional comparisons (not presented here due to space limitations) between SWC and SEC have further shown that the gains of SWC over SEC increase as $L$ decreases. 

It can be also observed from Fig$.$~\ref{Fig:ASEP_SWD_MRD} that, as $L$ decreases and/or $\rho$ increases, the $\overline{g}_1$ values for which
the ABEP of SWC becomes better than the ABEP of MRC decrease. On the other hand, as shown in Fig$.$~\ref{Fig:Optimum_gT_SWD_MRD}, by increasing $\overline{g}_1$ and/or $m$, and/or decreasing $\rho$, this increases $g_{{\rm T}_1}$ for the SWC reception. Furthermore, it can be observed that the impact of increasing $\rho$ is more severe to MRC than SWC and this difference becomes larger as $\overline{g}_1$ decreases. For example, for $\overline{P}_{\rm e}=10^{-3}$ with $2$- and $5$-branch MRC reception, the $\overline{g}_1$ gap between the considered IID and AC-NNID case is approximately $8.5$ dB whereas, for the case of SWC this gap decreases to approximately $3.5$ dB. For the ABEP results in Fig$.$~\ref{Fig:ASEP_SWD_MRD}, it can be seen that the AWT of $L$-branch SEC is independent of $\overline{g}_1$. Furthermore from the same figure, it can be seen that for the case of IID Rayleigh fading, $\overline{N}_{\rm c}=0.429$ and $0.487$ for $L=2$ and $5$, respectively, while for the AC-NNID case $\overline{N}_{\rm c}=0.423$ for $L=2$ and $\overline{N}_{\rm c}=0.527$ for $L=5$.   

The impact of the modulation order $M$ and $\rho$ on the ABEP performance of $L$-branch SWC receivers is demonstrated in Figs$.$~\ref{Fig:ASEP_SWD_MRD_MQAM} and~\ref{Fig:ASEP_vs_rho}. As expected, increasing $M$ degrades ABEP and, for a given $\overline{g}_1$ value, the impact of $\rho$ on ABEP increases with decreasing $M$. As observed in Fig$.$~\ref{Fig:ASEP_vs_rho} for low values of $\rho$ and decreasing $M$, ABEP improves when $m$ becomes large. Interestingly, as $\rho$ and $\overline{g}_1$ increase as well as $M$ decreases, ABEP degrades with increasing $m$.

\section{Conclusion} \label{Sec:Conclusion}
In this paper, we have analyzed and evaluated the performance of $L$-branch SWC receivers over AC-NNID Nakagami-$m$ fading channels. Fast convergent infinite series representations for the average error performance of various modulation schemes as well as the ANPE and AWT of the considered receivers were presented. As shown from the numerically evaluated results which have been verified by means of computer simulations, the superiority of error performance of SWC over SEC increases as fading conditions and correlated fading become more severe at the cost of negligible increasing AWT. It was also demonstrated that: \textit{i}) the impact of increasing correlation is more severe to the average error performance of MRC than that of SWC; and \textit{ii}) if the diversity reception system is resilient to some AWT, there are certain cases where the average error probability of SWC is superior to that of MRC.

\section*{Acknowledgment}
The work of P$.$ Fan was supported by $111$ Project (No$.$ $111$-$2$-$14$) and NSFC (No$.$ $61471302$).

\bibliographystyle{IEEEtran}
\bibliography{IEEEabrv,SWD_references}
\end{document}